\documentclass[a4paper,conference]{IEEEtran}

\IEEEsettopmargin{t}{30mm}
\IEEEquantizetextheight{c}
\IEEEsettextwidth{14mm}{14mm}
\IEEEsetsidemargin{c}{0mm}

\usepackage[utf8]{inputenc}
\usepackage[T1]{fontenc}
\usepackage{url}
\usepackage{ifthen}
\usepackage{cite}
\usepackage{xcolor}
\usepackage[cmex10]{amsmath} 
\usepackage{amssymb}
\usepackage{comment}

\newtheorem{theorem}{Theorem}
\newtheorem{lemma}{Lemma}

\usepackage{amsopn, graphicx} 
\newcommand{\set}[1]{\mathcal{#1}}
\newcommand{\I}[1]{\mathbf{1}\left\{#1\right\}}
\newcommand{\argmax}{\operatorname*{argmax}}
\newcommand{\argmin}{\operatorname*{argmin}}
\newcommand{\E}[1]{\mathsf{E}\left[#1\right]}

\begin{document}
\title{Universal Outlier Hypothesis Testing\\via Mean- and Median-Based Tests}



\author{%
  \IEEEauthorblockN{Bernhard C. Geiger\IEEEauthorrefmark{1}\IEEEauthorrefmark{2}\IEEEauthorrefmark{5},
  Tobias Koch\IEEEauthorrefmark{3}\IEEEauthorrefmark{4},
  Josipa Mihaljevi\'c\IEEEauthorrefmark{2}, and
  Maximilian Toller\IEEEauthorrefmark{2}}
  \IEEEauthorblockA{\IEEEauthorrefmark{1}%
             Signal Processing and Speech Communication Laboratory, Graz University of Technology, 8010 Graz, Austria}
  \IEEEauthorblockA{\IEEEauthorrefmark{2}%
             Know Center Research GmbH, 8010 Graz, Austria}

    \IEEEauthorblockA{\IEEEauthorrefmark{5}%
             Graz Center for Machine Learning, 8010 Graz, Austria}
  \IEEEauthorblockA{\IEEEauthorrefmark{3}%
             Signal Theory and Communications Department, Universidad Carlos III de Madrid, 28911 Legan\'es, Spain}
  \IEEEauthorblockA{\IEEEauthorrefmark{4}%
             Gregorio Mara\~n\'on Health Research Institute, 28007 Madrid, Spain}
  \IEEEauthorblockA{e-mails: geiger@ieee.org, $\{$jmihaljevic,mtoller$\}$@know-center.at, tkoch@ing.uc3m.es}
}

\maketitle
\begingroup\renewcommand\thefootnote{}
\footnotetext{Author names are listed in alphabetical order.}
\endgroup

\begin{abstract}
Universal outlier hypothesis testing refers to a hypothesis testing problem where one observes a large number of length-$n$ sequences---the majority of which are distributed according to the typical distribution $\pi$ and a small number are distributed according to the outlier distribution $\mu$---and one wishes to decide, which of these sequences are outliers without having knowledge of $\pi$ and $\mu$. In contrast to previous works, in this paper it is assumed that both the number of observation sequences and the number of outlier sequences grow with the sequence length. In this case, the typical distribution $\pi$ can be estimated by computing the mean over all observation sequences, provided that the number of outlier sequences is sublinear in the total number of sequences. It is demonstrated that, in this case, one can achieve the error exponent of the maximum likelihood test that has access to both $\pi$ and $\mu$. However, this mean-based test performs poorly when the number of outlier sequences is proportional to the total number of sequences. For this case, a median-based test is proposed that estimates $\pi$ as the median of all observation sequences. It is demonstrated that the median-based test achieves again the error exponent of the maximum likelihood test that has access to both $\pi$ and $\mu$, but only with probability approaching one. To formalize this case, the typical error exponent---similar to the typical random coding exponent introduced in the context of random coding for channel coding---is proposed.
\end{abstract}

\section{Introduction}
The problem of \emph{outlier hypothesis testing}, introduced by Li, Nitinawarat, and Veeravalli \cite{Li_Universal}, consists in deciding among a large number of  sequences which ones are outliers. More precisely, in outlier hypothesis testing, we observe $\mathsf{M}$ independent length-$n$ sequences $\mathbf{Y}_1,\ldots,\mathbf{Y}_{\mathsf{M}}$, $\mathsf{M}-\mathsf{T}$ of which are  independent and identically distributed (i.i.d.) according to the \emph{typical distribution $\pi$}, and $\mathsf{T}<\mathsf{M}/2$ are i.i.d.\ according to the \emph{outlier distribution  $\mu \neq \pi$}, both taking values from the same finite alphabet $\set{Y}$. The goal is to decide which sequences are distributed according to $\mu$, i.e., to identify the outlier sequences. For any fixed $\mathsf{M}$, this can be formulated as the problem of finding optimizers of a functional over the search space $[\mathsf{M}]\triangleq\{1,\dots,\mathsf{M}\}$.

We shall describe the outlier hypothesis test by the function $\delta\colon \mathcal{Y}^{\mathsf{M}n} \to \bar{\mathcal{S}}$, where $\bar{\mathcal{S}}$ denotes the set of size-$\mathsf{T}$ subsets of $[\mathsf{M}]$. In words, the function $\delta(\cdot)$ takes the $\mathsf{M}$ sequences $\mathbf{Y}_1,\ldots,\mathbf{Y}_{\mathsf{M}}$ as input and produces the set of indices $\mathcal{S}$ indicating the outlier sequences. The performance of this test is measured by the maximal error probability
\begin{equation}
\epsilon(\delta) = \max_{\mathcal{S}\in \bar{\mathcal{S}}} \sum_{\mathbf{y}_1^{\mathsf{M}}\colon \delta(\mathbf{y}_1^{\mathsf{M}}) \neq \mathcal{S}} P_{\mathcal{S}}(\mathbf{y}_1^{\mathsf{M}})
\end{equation}
where $P_{\mathcal{S}}$ denotes the joint distribution of $\mathbf{Y}_1,\ldots,\mathbf{Y}_{\mathsf{M}}$ when the outlier sequences have indices $\mathcal{S}$, and where we use the notation $\mathbf{y}_1^{\mathsf{M}}$ to denote the sequence $\mathbf{y}_1,\ldots,\mathbf{y}_{\mathsf{M}}$. More precisely, we study the decay at which $\epsilon(\delta)$ tends to zero as $n\to\infty$ by considering the corresponding error exponent 
\begin{equation}
    \alpha(\delta) \triangleq \varlimsup_{n\to\infty}-\frac{1}{n} \log \epsilon(\delta) \label{eq:error_exp_def}
\end{equation}
where $\log(\cdot)$ is the binary logarithm function and $\varlimsup$ denotes the \emph{limit superior}.

If both distributions $\pi$ and $\mu$, and the number $\mathsf{T}$ of outlier sequences are known, then the error exponent of the maximum likelihood test is $2B(\pi,\mu)$~\cite[Prop.~7]{Li_Universal}, where 
\begin{equation}
    B(\pi,\mu)\triangleq -\log\left(\sum_{y\in\set{Y}} \pi(y)^{\frac{1}{2}}\mu(y)^{\frac{1}{2}}\right)
\end{equation}
is the Bhattacharyya distance between $\pi$ and $\mu$. The same exponent can also be achieved by a test that has only access to $\pi$ but is ignorant of $\mu$ \cite[Th.~8]{Li_Universal}.

In the universal setting, where neither $\pi$ nor $\mu$ are known, the maximum likelihood test is inapplicable. Instead, Li, Nitinawarat, and Veeravalli consider the generalized likelihood ratio test (GLRT), which is given by \cite[eq.~(37)]{Li_Universal}
\begin{IEEEeqnarray}{lCl}
    \IEEEeqnarraymulticol{3}{l}{\delta_{\mathrm{Li}}(\mathbf{Y}_1,\ldots,\mathbf{Y}_{\mathsf{M}})} \nonumber\\
    \quad &= &  \argmin_{\set{S}\in\bar{\set{S}}} \sum_{j\notin\set{S}} D\left(\mathsf{P}_{\mathbf{Y}_j} \Bigg|\Bigg| \frac{ \sum_{k\notin\set{S}} \mathsf{P}_{\mathbf{Y}_k}}{\mathsf{M}-\mathsf{T}}\right) \label{eq:Li_GLRT}
\end{IEEEeqnarray}
where
\begin{equation}
\mathsf{P}_{\mathbf{Y}_k}(y) \triangleq \frac{1}{n} \sum_{\ell=1}^n \I{Y_{k,\ell} = y}, \quad y \in\mathcal{Y}
\end{equation}
denotes the empirical distribution (type) of \mbox{$\mathbf{Y}_k=(Y_{k,1},\ldots,Y_{k,n})$}, and $\I{\cdot}$ denotes the indicator function. For the GLRT, they derive a lower bound on the error exponent, which is strictly smaller than $2 B(\pi,\mu)$, but converges to $2 B(\pi,\mu)$ as the number of observation sequences $\mathsf{M}$ tends to infinity \cite[Th.~10]{Li_Universal}. This agrees with the intuition that, as the number of observation sequences tends to infinity, we can accurately estimate the typical distribution by averaging over all observations.

The work of Li, Nitinawarat, and Veeravalli \cite{Li_Universal} has subsequently been extended in various directions. For example, a test for $\mathsf{T}\le 1$ with rejection option, i.e., where the test can decide that none of the observation sequences is an outlier, was investigated in~\cite{Zhou_RejectOption}. Nitinawarat and Veeravalli \cite{Nitinawarat_Stopping} considered the case where in exactly one sequence a change point occurs at time $1\le\lambda\le n$ and proposed a test for identifying the sequence and $\lambda$. Bu \emph{et al.} \cite{Bu_Continuous} proposed tests for continuous distributions $\pi$ and $\mu$. Sequential detection, i.e., early stopping, was considered in~\cite{Li_Sequential_ISIT,Li_Sequential_Asilomar,Sreenivasan_DataStreams}. Acharya \emph{et al.} \cite{Acharya_Sublinear} studied the setting where the sample size $n$ is smaller than the alphabet size $|\set{Y|}$ of the discrete distributions.

The aforementioned works all concern the case where the sample size $n$ is significantly larger than the number $\mathsf{M}$ of observation sequences. However, sometimes the opposite case is also of interest. A practical example are dense sensor networks, where each of the $\mathsf{M}$ sensors can only sense over short intervals or with low sampling rates due to energy limitations. Another example occurs in the medical domain, where a large number of patients are recorded for a comparably short time, such as for ECG/EEG data or voice recordings. A large number of observation sequences has the advantage that the typical distribution can be estimated accurately as
\begin{equation}
\label{eq:pihat}
\hat{\pi}(y) = \frac{1}{\mathsf{M}} \sum_{k=1}^{\mathsf{M}} \mathsf{P}_{\mathbf{Y}_k}(y), \quad y\in\mathcal{Y}
\end{equation}
provided that the number $\mathsf{T}$ of outlier sequences is sublinear in $\mathsf{M}$. On the negative side, the GLRT may become infeasible if the number of observation sequences is large. Indeed, the test searches over all size-$\mathsf{T}$ subsets of $[\mathsf{M}]$ as candidate outlier sets. It thus has a computational complexity of $\mathcal{O}(\mathsf{M}^\mathsf{T})$, which becomes prohibitive if $\mathsf{M}$ and $\mathsf{T}$ are large. Mathematically, we model the case of a large number of observation sequences by making the parameters $\mathsf{M}$ (number of observation sequences) and $\mathsf{T}$ (number of outlier sequences) dependent on the sequence length $n$. As a consequence, the approach followed by Li, Nitinawarat, and Veeravalli \cite{Li_Universal} of applying Sanov's theorem to express the error exponent as a minimization of relative entropies and then bounding this minimum cannot be followed for the GLRT, since the subexponential terms of the error probability depend exponentially on the alphabet size $\mathcal{Y}^{\mathsf{M}}$ of the sequences $\mathbf{Y}_1,\ldots,\mathbf{Y}_{\mathsf{M}}$ and may become significant for an $n$-dependent $\mathsf{M}$.

To sidestep this problem, we propose the \emph{mean-based test}
\begin{equation}\label{eq:mean_test}
    \delta_{\mathrm{mean}}(\mathbf{Y}_1,\ldots,\mathbf{Y}_{\mathsf{M}}) = \argmax_{\set{S}\in \bar{\set{S}}} \sum_{j\in\set{S}} D\left(\mathsf{P}_{\mathbf{Y}_j} \big\| \hat{\pi} \right)
\end{equation}
whose error probability, conditioned on $\hat{\pi}$, can be analyzed analytically. In particular, we show that, if $\mathsf{T}=o(\mathsf{M})$ and $\mathsf{M}$ grows superlinearly but subexponentially in $n$, then the mean-based test achieves the optimal error exponent $2B(\pi,\mu)$.

If the outlier sequences represent a positive fraction of all sequences, i.e., if $\mathsf{T}=c\mathsf{M}$, $0<c<\frac{1}{2}$, then the mean over all types is biased, which negatively affects the error exponent of the mean-based test \eqref{eq:mean_test}. In this case, the median over all types
\begin{IEEEeqnarray}{lCl}
    \overline{\pi}(y) = \frac{\operatorname{median}\{\mathsf{P}_{\mathbf{Y}_1}(y),\dots,\mathsf{P}_{\mathbf{Y}_\mathsf{M}}(y)\}}{\sum_{y'\in\set{Y}}\operatorname{median}\{\mathsf{P}_{\mathbf{Y}_1}(y'),\dots,\mathsf{P}_{\mathbf{Y}_\mathsf{M}}(y')\}},\quad\nonumber\\
    \IEEEeqnarraymulticol{3}{r}{ \quad y\in\mathcal{Y}}\label{eq:pi_est}
\end{IEEEeqnarray}
may provide a better estimate of $\pi$, since it is more robust against outliers. As we shall show, the median becomes accurate as the sequence length $n$ tends to infinity but, in contrast to the mean, its accuracy does not improve with $\mathsf{M}$. This implies that the error exponent of a median-based test that replaces in \eqref{eq:mean_test} the mean-based estimate $\hat{\pi}$ by the median-based estimate $\bar{\pi}$ is poor, despite the robustness of the median against outliers, and despite numerical results that seem to suggest otherwise. We thus propose a two-step median-based test that computes the median-based estimate $\bar{\pi}$ from a part of the sequences $\mathbf{Y}_1,\ldots,\mathbf{Y}_{\mathsf{M}}$ and uses the remaining part for outlier testing. As a less pessimistic performance measure, we then propose the \emph{typical} error exponent, defined as the average error exponent averaged over all realizations of $\bar{\pi}$. The typical error exponent is similar to the typical random coding exponent introduced in the context of random coding for channel coding; see, e.g., \cite{Barg02,Nazari14,Merhav18,Truong24}. We show that the typical error exponent of the two-step median-based test is equal to the optimal error exponent $2B(\pi,\mu)$. Thus, the test achieves the optimal exponent with probability approaching one.

In terms of complexity, the mean-based test and the two-step median-based test have a computational complexity of $\set{O}(\mathsf{M}\log \mathsf{M})$, while searching for a size-$c\mathsf{M}$ subset, as in the GLRT \eqref{eq:Li_GLRT}, has complexity $\mathcal{O}(2^{2\mathsf{M}}/\sqrt{\mathsf{M}})$.

The rest of this paper is organized as follows. Section~\ref{sec:setup} introduces the considered setup. Section~\ref{sec:mean_test} discusses the error exponent of the mean-based test. Section~\ref{sec:median_test} studies the two-step median-based test and presents its typical error exponent. Section~\ref{sec:numerical} compares the performances of the proposed tests with that of the GLRT by means of numerical examples. Section~\ref{sec:conclusions} concludes the paper with a discussion of the computation complexity of the considered tests.

\section{Setup}
\label{sec:setup}
Suppose we observe $\mathsf{M}_n$ independent length-$n$ sequences $\mathbf{Y}_1,\ldots,\mathbf{Y}_{\mathsf{M}_n}$, $\mathsf{M}_n-\mathsf{T}_n$ of which are i.i.d.\ according to the typical distribution $\pi$ and $\mathsf{T}_n<\mathsf{M}_n/2$ are i.i.d.\ according to the outlier distribution  $\mu \neq \pi$. Here and throughout the rest of this paper, we add the subscript ``$n$'' to $\mathsf{M}$ and $\mathsf{T}$ to reflect in the notation that these two parameters may depend on the sequence length $n$. We assume that  $\mu$ and $\pi$ have the same alphabet $\set{Y}$, which is assumed to be finite. We further assume that
\begin{equation}
    \pi_{\min} \triangleq \min_{y\in\set{Y}} \pi(y) >0.
\end{equation}
By the symmetry of the problem, we can assume without loss of generality that the first $\mathsf{T}_n$ sequences are outliers. Under this assumption, the sequences $\mathbf{Y}_1,\ldots,\mathbf{Y}_{\mathsf{M}_n}$ have the joint distribution
\begin{equation}
P_{[\mathsf{T}_n]}(\mathbf{y}_1,\dots,\mathbf{y}_{\mathsf{M}_n})
=  \prod_{k=1}^n\prod_{i=1}^{\mathsf{T}_n} \mu(y_{i,k}) \prod_{j=\mathsf{T}_n+1}^{\mathsf{M}_n} \pi(y_{j,k}) \label{eq:P_T}
\end{equation}
where $\mathbf{y}_j=(y_{j,1},\dots,y_{j,n})^T$, $j=1,\dots,\mathsf{M}_n$. We shall assume that the number of outlier sequences $\mathsf{T}_n$ is known. Any outlier test can thus be written as a function $\delta(\mathbf{Y}_1,\ldots,\mathbf{Y}_{\mathsf{M}_n})\colon \set{Y}^{n\mathsf{M}_n}\to \bar{\set{S}}$, where $\bar{\mathcal{S}}$ denotes the set of size-$\mathsf{T}_n$ subsets of $[\mathsf{M}_n]$, i.e., the set of all sets of cardinality $\mathsf{T}_n$ in the powerset $2^{[\mathsf{M}_n]}$. The maximal error probability can then be written as
\begin{equation}
    \epsilon(\delta) = \sum_{\mathbf{y}_1^{\mathsf{M}_n}\colon \delta(\mathbf{y}_1^{\mathsf{M}_n})\neq [\mathsf{T}_n]} P_{[\mathsf{T}_n]}(\mathbf{y}_1^{\mathsf{M}_n})
\end{equation}
and the corresponding error exponent is as in \eqref{eq:error_exp_def}.

For the two-step median-based test, we further introduce the \emph{typical error exponent}. Specifically, suppose that we use the first $\lceil \rho n \rceil$, $\rho\in(0,1)$ samples of each one of the vectors $\mathbf{Y}_1,\ldots,\mathbf{Y}_{\mathsf{M}_n}$, denoted by  $\mathbf{Y}_1^{(1)},\ldots,\mathbf{Y}_{\mathsf{M}_n}^{(1)}$, to produce the median-based estimate $\bar{\pi}$ of $\pi$ (cf.~\eqref{eq:pi_est}) and the remaining samples $\mathbf{Y}_1^{(2)},\ldots,\mathbf{Y}_{\mathsf{M}_n}^{(2)}$ to decide which sequences are outliers. Thus, in the second step, the outlier test is given by the function
 \begin{equation}
 \delta_{\textnormal{median}}(\mathbf{Y}_1^{(2)},\ldots,\mathbf{Y}_{\mathsf{M}_n}^{(2)} | \bar{\pi}) = \argmax_{\set{S}\in \bar{\set{S}}} \sum_{j\in\set{S}} D\left(\mathsf{P}_{\mathbf{Y}_j^{(2)}} \big\| \bar{\pi} \right).\label{eq:median_test}
 \end{equation}
 The maximal error probability in the second step can be written as
 \begin{equation}
    \epsilon(\delta_{\textnormal{median}} | \bar{\pi}) = \sum_{\mathbf{y}_1^{\mathsf{M}_n}\colon \delta_{\textnormal{median}}(\mathbf{y}_1^{\mathsf{M}_n} | \bar{\pi})\neq [\mathsf{T}_n]} P^{(2)}_{[\mathsf{T}_n]}(\mathbf{y}_1^{\mathsf{M}_n})
\end{equation}
where $P^{(2)}_{[\mathsf{T}_n]}$ is as in \eqref{eq:P_T} but with the index $k$ going from $\lceil \rho n \rceil +1$ to $n$. The typical error exponent is then defined as
 \begin{equation}
\bar{\alpha}(\delta_{\textnormal{median}}) \triangleq \varlimsup_{n\to\infty} -\frac{1}{n}\mathsf{E}\left[\log\bigl(\epsilon(\delta_{\textnormal{median}}|\bar{\pi})\bigr)\right] \label{eq:typical_exp}
 \end{equation}
 where the expectation is over all observation sequences $\mathbf{Y}_1^{(1)},\ldots,\mathbf{Y}_{\mathsf{M}_n}^{(1)}$ used in step 1 to produce $\bar{\pi}$. 

Jensen's inequality implies that the typical error exponent $\bar{\alpha}(\delta_{\textnormal{median}})$ is never smaller than the error exponent $\alpha(\delta_{\textnormal{median}})$. Intuitively, to obtain the optimal typical error exponent, it suffices that the estimate $\bar{\pi}$ is accurate with probability approaching one. In contrast, the error exponent requires the probability that $\bar{\pi}$ deviates from $\pi$ to decay exponentially in $n$ with a sufficiently large error exponent. Our numerical results in Section~\ref{sec:numerical} suggest that such a requirement may be too pessimistic. 

\section{Mean-Based Test}
\label{sec:mean_test}
We start by investigating the mean-based test in~\eqref{eq:mean_test}. The following theorem characterizes the performance of this test when the number of outlier sequences is sublinear in $\mathsf{M}_n$.

\begin{theorem}\label{thm:mean_test}
    Assume the number of sequences $\mathsf{M}_n$ satisfies $\lim_{n\to\infty} \mathsf{M}_n/n=\infty$ and $\lim_{n\to\infty} \log(\mathsf{M}_n)/n =0$. Further assume that the number of outlier sequences is known and satisfies $\mathsf{T}_n=o(\mathsf{M}_n)$. Then, for every pair of distributions $\mu\neq\pi$, the mean-based test \eqref{eq:mean_test} has the error exponent
    \begin{equation}
        \alpha_\mathrm{mean} = 2B(\pi,\mu).
    \end{equation}
\end{theorem}

\begin{IEEEproof}
    See Appendix~\ref{proof:thm:mean_test}.
\end{IEEEproof}

Theorem~\ref{thm:mean_test} hinges on the assumption that $\mathsf{T}_n$ is sublinear in $\mathsf{M}_n$, so the estimate $\hat{\pi}$ of $\pi$ becomes accurate as $n\to\infty$. When $\mathsf{T}_n$ is proportional to $\mathsf{M}_n$, this is no longer the case. In particular, if $\mathsf{T}_n=c\mathsf{M}_n$, $c>0$, then $\hat{\pi}$ converges to the distribution $\nu=(1-c)\pi + c\mu$. By following the same steps as in the proof of Theorem~\ref{thm:mean_test}, it can be shown that the error exponent is lower-bounded by
\begin{subequations}\label{eq:mean_test:cM}
        \begin{equation}
            \alpha_\mathrm{mean} = \min_{Q_1,Q_2} \left\{D(Q_1\|\mu) + D(Q_2\| \pi)\right\}
        \end{equation}
        where the minimum is over all distributions $(Q_1,Q_2)$ satisfying
        \begin{equation}
            D(Q_2|\nu) - D(Q_1 \| \nu) \geq 0
        \end{equation}
    \end{subequations}
which is strictly smaller than $2B(\pi,\mu)$, cf.~Fig.~\ref{fig:exp2}. Thus, when the number of outlier sequences is proportional to the total number of sequences $\mathsf{M}_n$, the mean-based test does not achieve the exponent of the maximum-likelihood test when $\pi$ and $\mu$ are known.

\section{Median-Based Test}
\label{sec:median_test}

Since the mean-based test performs subpar for $\mathsf{T}_n=c\mathsf{M}_n$, we consider in this section the median-based estimate $\bar{\pi}$ (cf.~\eqref{eq:pi_est}), which is inherently robust against outliers as long as $c<\frac{1}{2}$. However, in contrast to the mean-based estimate $\hat{\pi}$, the convergence of the median-based estimate \eqref{eq:pi_est} to $\pi$ is determined by the growth of $n$ and not of $\mathsf{M}_n$. As a consequence, the probability that $\bar{\pi}\notin \set{B}_{\varepsilon,\infty}(\pi)$, where
\begin{IEEEeqnarray}{lCl}
\IEEEeqnarraymulticol{3}{l}{\set{B}_{\varepsilon,\infty}(\pi)} \nonumber\\
\quad & \triangleq & \{Q \in \set{P}(\set{Y})\colon Q(y) \in [\pi(y)-\varepsilon,\pi(y)+\varepsilon], y\in\set{Y}\}\IEEEeqnarraynumspace
\end{IEEEeqnarray}
does not decay superexponentially in $n$, even if $\mathsf{M}_n$ is superlinear in $n$. In fact, this probability can be bounded as follows:

\begin{lemma}\label{lemma:median_conv}
Consider the median-based estimate $\bar{\pi}$ of $\pi$, as given in \eqref{eq:pi_est}, and assume that $\mathsf{T}_n < \mathsf{M}_n/2$. Then, for every $\varepsilon>0$,
\begin{equation}
P_{[\mathsf{T}_n]}\bigl(\bar{\pi}\notin \set{B}_{\varepsilon,\infty}(\pi)\bigr) \leq 2 |\set{Y}|(\mathsf{M}_n-\mathsf{T}_n) \exp(-2 n \varepsilon'^2) \label{eq:lemma_median_statement}
\end{equation}
where $\varepsilon' = \varepsilon/(1+|\set{Y}|)$.
\end{lemma}

\begin{IEEEproof}
    See Appendix~\ref{proof:lemma:median_conv}.
\end{IEEEproof}

Observe that the RHS of \eqref{eq:lemma_median_statement} decays exponentially in $n$, but the error exponent vanishes as $\varepsilon'$ tends to zero. Thus, if we reproduce the steps in the proof of Theorem~\ref{thm:mean_test} for the median-based estimate, then we obtain a vanishing error exponent. This contradicts the numerical results in Section~\ref{sec:numerical}, which demonstrate that, for a finite $n$, the median-based test outperforms the mean-based test; cf.~Fig.~\ref{fig:exp3}. One possible explanation for this mismatch is that the error exponent is too pessimistic, since it is dominated by the slow convergence of $\bar{\pi}$ to $\pi$. This motivates us to study the typical error exponent, defined in \eqref{eq:typical_exp}, instead.

To simplify the analysis, we shall characterize the typical error exponent of the two-step median-based test described at the end of Section~\ref{sec:setup}. Specifically, we split each sequence $\mathbf{Y}_i$ into two subsequences $\mathbf{Y}_i^{(1)}$ and $\mathbf{Y}_i^{(2)}$ that consist of the first $\lceil \rho n\rceil$ and the remaining $n-\lceil \rho n\rceil$ samples of $\mathbf{Y}_i$, respectively. The former subsequences are then used to estimate $\pi$, while the latter subsequences are used to detect the outliers. Since the observation sequences are i.i.d., the estimate $\bar{\pi}$ will then be independent of the subsequences $\mathbf{Y}_1^{(2)},\ldots,\mathbf{Y}_{\mathsf{M}_n}^{(2)}$, which is easier to analyze. The following theorem characterizes the typical error exponent of this test.
\begin{theorem}\label{thm:median_test}
 Consider the two-step median-based test described at the end of Section~\ref{sec:setup}, and assume that $\lim_{n\to\infty} \log(\mathsf{M}_n)/n =0$. Further assume that the number of outlier sequences is known and satisfies $\mathsf{T}_n<\mathsf{M}_n/2$. Then, for every pair of distributions $\mu\neq\pi$, the test~\eqref{eq:median_test} has the typical error exponent
 \begin{equation}
     \bar{\alpha}_\mathrm{median} = 2B(\pi,\mu).
 \end{equation}
\end{theorem}
\begin{IEEEproof}
See Appendix~\ref{proof:thm:median_test}.
\end{IEEEproof}

Theorem~\ref{thm:median_test} only requires that $\mathsf{M}_n$ is subexponential in $n$. Thus, in contrast to Theorem~\ref{thm:mean_test}, the theorem also applies to the case where $\mathsf{M}_n$ does not grow with $n$. As such, we can compare the typical error exponent obtained in Theorem~\ref{thm:median_test} with the results obtained in \cite{Li_Universal}. Interestingly, the \emph{typical} error exponent is equal to the exponent of the maximum-likelihood test when $\pi$ and $\mu$ are known, whereas the error exponent is strictly smaller \cite[Th.~9]{Li_Universal}.

\section{Numerical Results}
\label{sec:numerical}

To validate our theoretical findings, we conduct three experiments:
\begin{enumerate}
    \item Numerical demonstration that the error exponent of the GLRT~\eqref{eq:Li_GLRT} is indeed often smaller than the typical error exponent of the two-step median-based test (Theorem~\ref{thm:median_test}).
    \item Numerical evaluation of~\eqref{eq:mean_test:cM} to confirm the suboptimality of the mean-based test when $\mathsf{T}_n=c\mathsf{M}_n$.
    \item Comparison of the mean-based test, the single-step median-based test,\footnote{The single-step median-based test uses the same observation sequences to produce the median-based estimate $\bar{\pi}$ of $\pi$ and to detect the outlier sequences.} and the two-step median-based test for increasing $c$ in $\mathsf{T}_n=c\mathsf{M}_n$ and increasing $\mathsf{M}_n$.
\end{enumerate}

\begin{figure}[t]
    \centering
    \includegraphics[width=\linewidth]{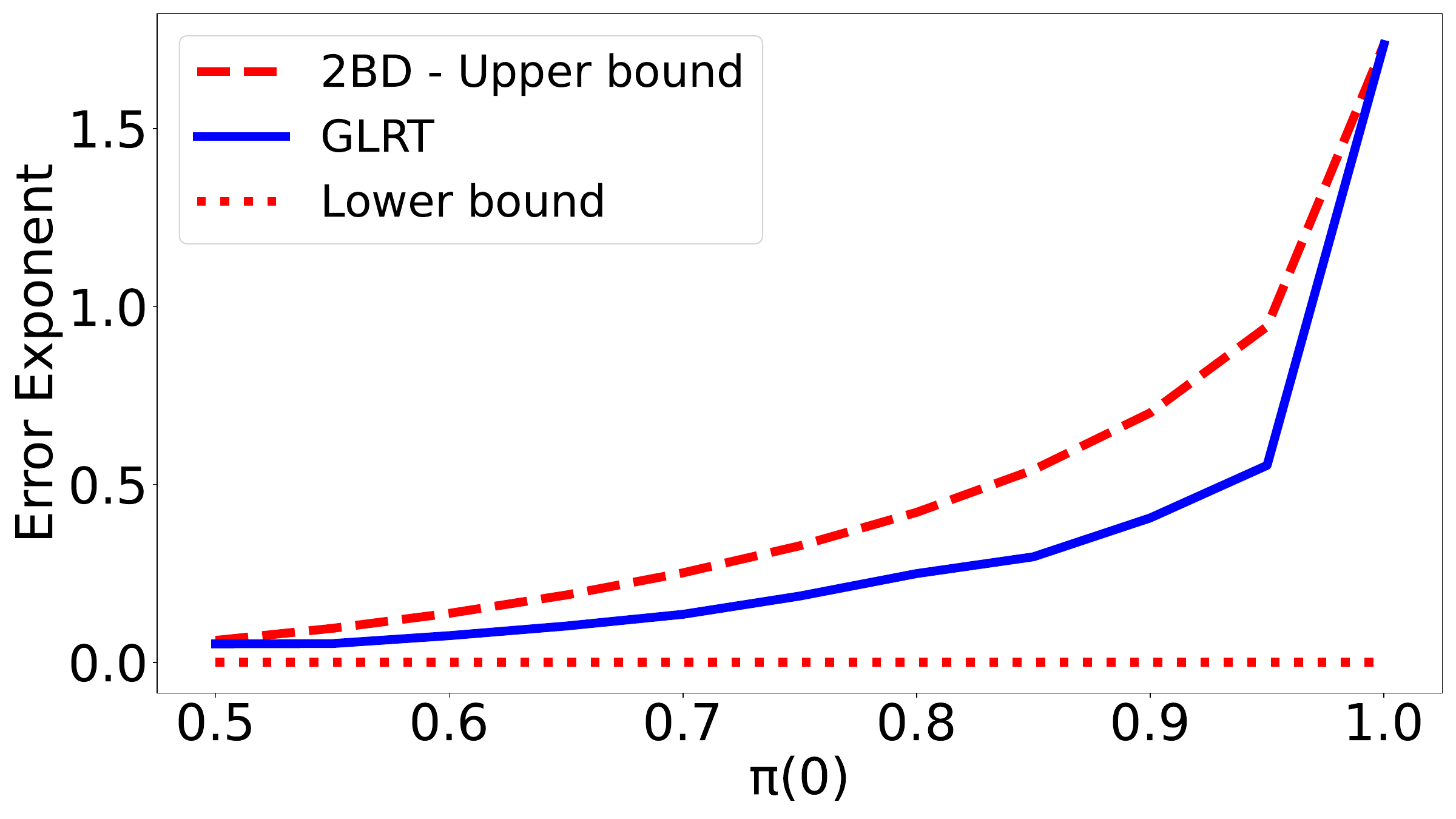}
    \caption{Numerical demonstration that the GLRT's error exponent is generally smaller than $2B(\pi,\mu)$.}
    \label{fig:exp1}
\end{figure}

\subsection{Suboptimality of GLRT} We select two Bernoulli distributions $\mu= B(0.7)$ and $\pi= B(\theta_\pi)$, and vary the probability $\pi(0)=1-\theta_\pi$ in the set $\{0.50,0.55,\dots,1.00\}$.
We select $\mathsf{M}_n=4$ and numerically approximate the error exponent of the GLRT \cite[Th.~2]{Li_Universal} and its lower bound \cite[Th.~3]{Li_Universal}. We solve the constituting optimization problems approximately by evaluating the objective functions and constraints on a grid. More specifically, for each $\pi(0)$ we evaluate the Bernoulli distributions $q_1$ through $q_\mathsf{4}$ in~\cite[eqs.~(20)--(21)]{Li_Universal} for success probabilities $\theta_{q_i}$, $i=1,\dots,4$ drawn from the set $\{0.00,0.05,\dots,1\}$. We proceed along similar lines to evaluate the lower bound~\cite[Th.~3]{Li_Universal}.
The results of this simulation are depicted in Fig.~\ref{fig:exp1}.
We observe that the GLRT's error exponent reaches $2B(\pi,\mu)$ only for \mbox{$\pi(0)= 0.5$} and $\pi(0) = 1$. While the performance of the GLRT will be better for larger $\mathsf{M}_n$ as per~\cite[Th.~3]{Li_Universal}, note that the lower bound from this theorem becomes vacuous for small values of $\mathsf{M}_n$. Indeed, for $\mathsf{M}_n=4$ our approximation of the lower bound evaluates to zero; cf.~\cite[Fig.~1]{Li_Universal}.

\begin{figure}[t]
    \centering
    \includegraphics[width=\linewidth]{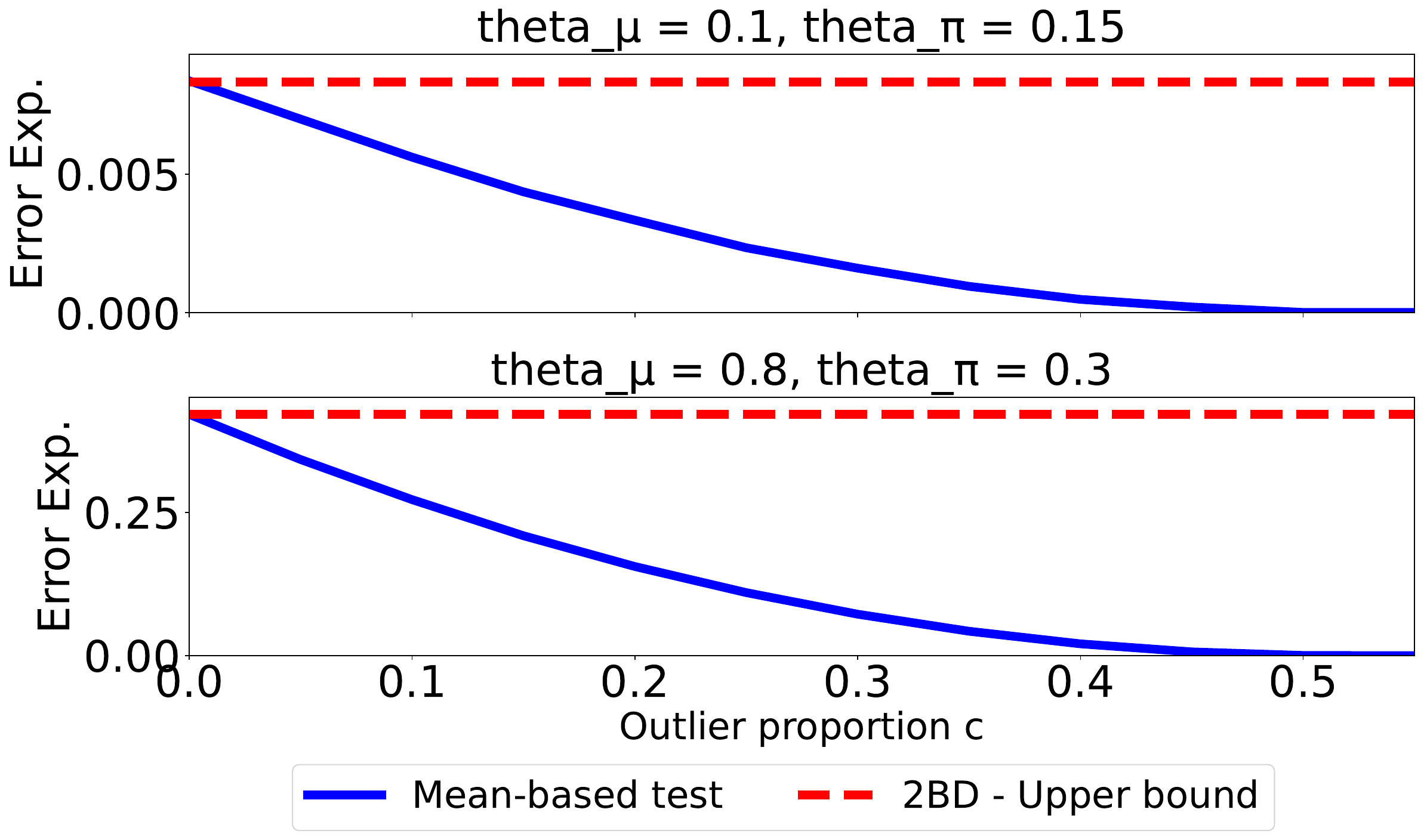}
    \caption{Error exponent $\alpha_{\text{mean}}$ of the mean-based test for increasing outlier proportion $c$ in two settings.}
    \label{fig:exp2}
\end{figure}

\subsection{Mean-based test when ${\mathsf{T}_n=c\mathsf{M}_n}$} With a positive outlier proportion $c$, the set of the observations $\mathbf{Y}_1^{\mathsf{M}_n}$ is drawn from the mixture distribution \mbox{$\nu = c \mu + (1-c) \pi$}.
We evaluate the corresponding error exponent of the mean-based test \eqref{eq:mean_test} for Bernoulli distributions (i.e., $\set{Y}=\{0,1\}$) and for outlier proportions $c$ in the set $\{0.00,0.05,\dots,0.55\}$.
We select the success probabilities $\theta_\mu$ and $\theta_\pi$ of the outlier and typical distribution, respectively, from the pairs $(0.10,0.15)$ and $(0.8,0.3)$. Hence, we evaluate the cases where these two distributions are similar and non-similar, respectively. We evaluate~\eqref{eq:mean_test:cM} numerically for Bernoulli distributions $Q_1$ and $Q_2$ assuming all success probabilities in the set $\{0.001,0.006,0.011,\dots,0.996\}$. 
Fig.~\ref{fig:exp2} shows the results for both pairs of $\theta_\mu,\theta_\pi$.
As expected, the error exponent is larger for non-similar distributions $\pi$ and $\mu$, and the error exponent of the mean-based test coincides with the upper bound of $2B(\pi,\mu)$ only if $c=0$. (Note that in this case we assume that $\mathsf{M}_n$ grows superlinearly with $n$.) We also observe that $\alpha_{\text{mean}}$ decays as $c$ increases and is zero if $c\ge0.5$, as in this case the mean $\hat{\pi}$ becomes a better estimate of the outlier distribution $\mu$ than of the typical distribution $\pi$.

\begin{figure}[t]
    \centering
    \includegraphics[width=\linewidth]{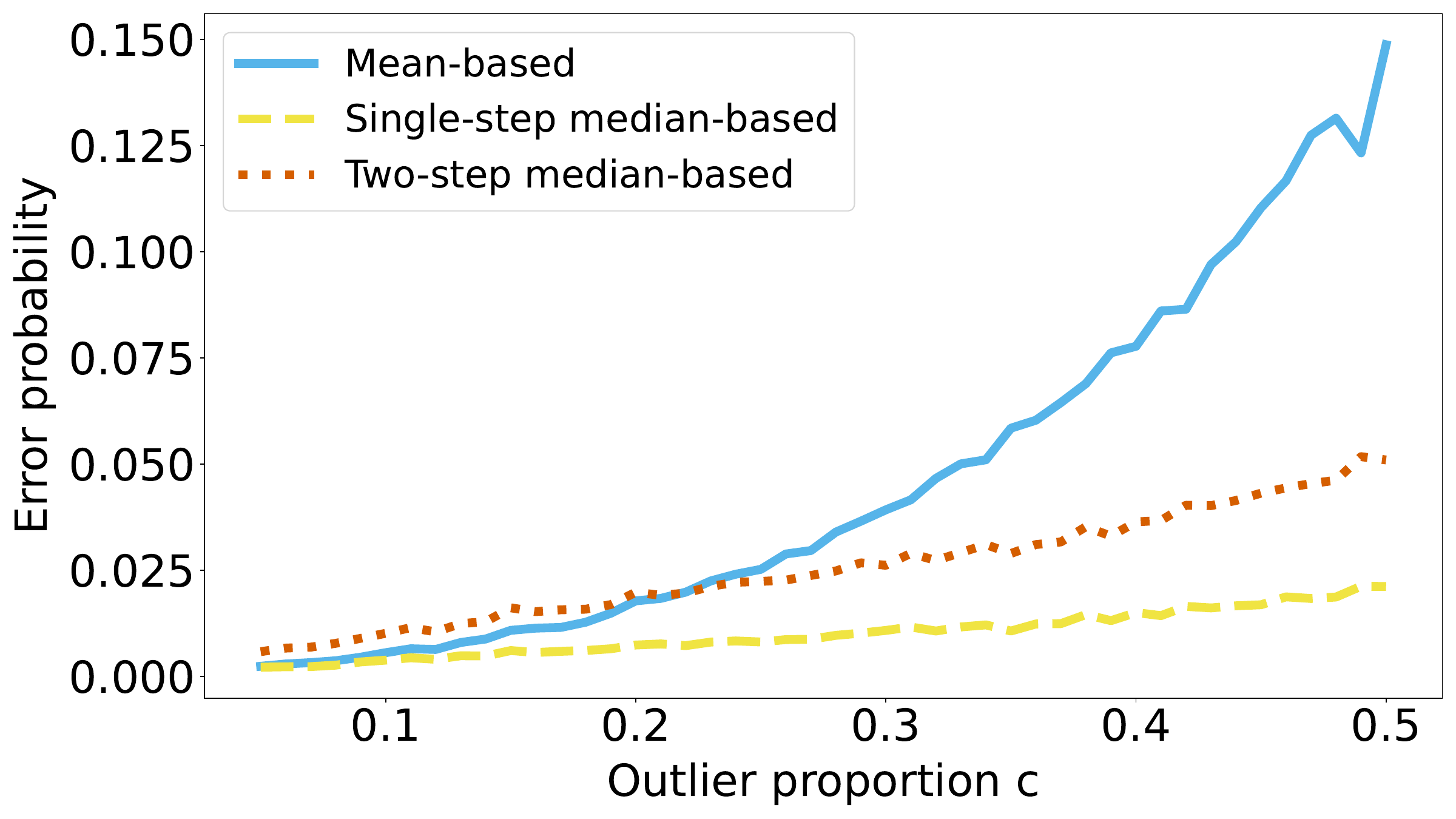}
    \caption{Comparison of mean-based, single-step median-based, and two-step median based tests over increasing outlier proportions $c$.}
    \label{fig:exp3}
\end{figure}

\subsection{Non-asymptotic performance of different tests} We finally evaluate the performance of the mean-based test, and the one-step and two-step median-based tests in the non-asymptotic regime, i.e., for finite $\mathsf{M}_n$ and $n$. Specifically, we set $\mathsf{M}_n=500$ and vary the outlier proportion $c$ in the set $\{0.05,0.06,\dots,0.5\}$. For each $c$, we select $\mu$ and $\pi$ as categorical distributions with $|\set{Y}|=5$ and with the individual symbol probabilities drawn from a uniform distribution and subsequently normalized to $1$. We next sample $\mathsf{T}_n=\lfloor c\mathsf{M}_n\rfloor$ and $\mathsf{M}_n-\mathsf{T}_n$ sequences of length $n=250$  from $\mu$ and $\pi$, respectively. From these, we compute the test statistics of the mean-based, single-step median-based, and two-step median-based tests, setting $\rho=0.5$ for the latter. Fig.~\ref{fig:exp3} shows the average error probability of each test, averaged over 200 independent runs of our experiments.
We observe that the mean-based test performs worst, unless $c$ is small, which is probably due to the biased estimate of $\pi$ when $\mathsf{T}_n$ is proportional to $\mathsf{M}_n$ (cf.~Fig.~\ref{fig:exp2}). Furthermore, the single-step median-based test generally outperforms the two-step test that we analyzed in Theorem~\ref{thm:median_test}. We believe that this is because the two-step test suffers from poor sample efficiency. However, for the one-step median-based test, we have not been able to obtain an expression for the typical error exponent, since in this case $\bar{\pi}$ and the observation sequences $\mathbf{Y}_1,\ldots,\mathbf{Y}_{\mathsf{M}_n}$, based on which we detect outliers, are dependent.

\section{Computational Complexity}
\label{sec:conclusions}

The mean-based and two-step median-based tests declare those $\mathsf{T}_n$ sequences as outliers for which the respective test statistics $D\left(\mathsf{P}_{\mathbf{Y}_j} \big\| \hat{\pi} \right)$ and $D\left(\mathsf{P}_{\mathbf{Y}_i^{(2)}} \big\| \bar{\pi} \right)$ are the largest. The tests hence require computing the type of each sequence (with a computational complexity of $\set{O}(n|\set{Y}|)$), taking the average or median over all types (computation complexity: $\set{O}(\mathsf{M}_n|\set{Y}|)$ or $\set{O}(|\set{Y}|\mathsf{M}_n\log\mathsf{M}_n)$, respectively), computing the test statistic for each of the $\mathsf{M}_n$ sequences (computational complexity: $\set{O}(\mathsf{M}_n|\set{Y}|)$), and sorting the $\mathsf{M}_n$ test statistics (computational complexity: $\set{O}(\mathsf{M}_n\log\mathsf{M}_n)$) to determine the $\mathsf{T}_n$ largest ones. Keeping $|\set{Y}|$ fixed, the optimal error exponent of $2B(\pi,\mu)$ can thus be achieved with a complexity of $\set{O}(n)+\set{O}(\mathsf{M}_n\log\mathsf{M}_n)$. In contrast, the GLRT proposed in \cite{Li_Universal} achieves this error exponent only asymptotically as $\mathsf{M}_n\to\infty$, but has a computational complexity of $\mathcal{O}(\mathsf{M}_n^{\mathsf{T}_n})$. Framing the test as a combinatorial clustering problem~\cite[Sec.~4.2]{Li_Clustering}, and approaching it suboptimally via K-means~\cite{Bu_LinearComplexity,Bu_LinearComplexityLong}, reduces the computational complexity to $\mathcal{O}(\mathsf{M}_n)$, but with an error exponent that is smaller than $2B(\pi,\mu)$~\cite[Th.~1~\&~3]{Bu_LinearComplexityLong}.

\appendices

\section{Proof of Theorem~\ref{thm:mean_test}}\label{proof:thm:mean_test}
Since the error exponent in the universal setting cannot be larger than the error exponent $2B(\pi,\mu)$ in the setting where $\pi$ and $\mu$ are known, it suffices to bound the error exponent from below.

We first show that, by the assumption that $\mathsf{T}_n = o(\mathsf{M}_n)$, the probability that the estimate $\hat{\pi}$ differs from $\pi$ by more than an $\varepsilon>0$ decays exponentially in $\mathsf{M}_n$. Since, by the assumption that $\lim_{n\to\infty} \mathsf{M}_n/n=\infty$, this then implies that this probability decays superexponentially in $n$. Indeed, defining
\begin{IEEEeqnarray}{lCl}
\IEEEeqnarraymulticol{3}{l}{\set{B}_{\varepsilon,\infty}(\pi)} \nonumber\\
\quad & \triangleq & \{Q \in \set{P}(\set{Y})\colon Q(y) \in [\pi(y)-\varepsilon,\pi(y)+\varepsilon], y\in\set{Y}\}\IEEEeqnarraynumspace
\end{IEEEeqnarray}
we have
\begin{IEEEeqnarray}{lCl}
    \IEEEeqnarraymulticol{3}{l}{P_{[\mathsf{T}_n]}(\hat{\pi}\notin \set{B}_{\varepsilon,\infty}(\pi))} \nonumber\\
  \quad &\leq & \sum_{y\in\set{Y}} P_{[\mathsf{T}_n]}\left(\Biggl|\pi(y)-\frac{1}{\mathsf{M}_n}\sum_{\ell=1}^{\mathsf{M}_n}\mathsf{P}_{\mathbf{Y}_{\ell}}(y)\Biggr| > \varepsilon\right)\nonumber\\
  \quad & = & \sum_{y\in\set{Y}} P_{[\mathsf{T}_n]}\left(\Biggl|\sum_{\ell=1}^{\mathsf{M}_n} \left(\mathsf{P}_{\mathbf{Y}_{\ell}}(y)-\pi(y)\right)\Biggr| > \varepsilon\mathsf{M}_n\right)
\end{IEEEeqnarray}
where the inequality follows from the union bound and the definitions of $\set{B}_{\varepsilon,\infty}(\pi)$ and $\hat{\pi}$. We next note that, under $P_{[\mathsf{T}_n]}$, 
\begin{equation}\label{eq:ExpecPT}
\mathsf{E}\left[\mathsf{P}_{\mathbf{Y}_{\ell}}(y)\right] = \begin{cases}\mu(y), \quad \ell=1,\dots,\mathsf{T}_n \\ \pi(y), \quad \ell=\mathsf{T}_n+1,\dots,\mathsf{M}_n\end{cases}
\end{equation}
for $y\in\set{Y}$. We can thus write
\begin{IEEEeqnarray}{lCl}
    \IEEEeqnarraymulticol{3}{l}{\Biggl|\sum_{\ell=1}^{\mathsf{M}_n} \left(\mathsf{P}_{\mathbf{Y}_{\ell}}(y)-\pi(y)\right)\Biggr|} \nonumber\\
  \quad &=& \Biggl|\sum_{\ell=1}^{\mathsf{M}_n} \left(\mathsf{P}_{\mathbf{Y}_{\ell}}(y)-\E{\mathsf{P}_{\mathbf{Y}_{\ell}}(y)}\right) +\mathsf{T}_n \mu(y) - \mathsf{T}_n\pi(y)\Biggr|\nonumber\\
  \quad &\leq&\Biggl|\sum_{\ell=1}^{\mathsf{M}_n} \left(\mathsf{P}_{\mathbf{Y}_{\ell}}(y)-\E{\mathsf{P}_{\mathbf{Y}_{\ell}}(y)}\right)\Biggr| +\mathsf{T}_n |\mu(y) - \pi(y)|\nonumber\\
  \quad &\leq&\Biggl|\sum_{\ell=1}^{\mathsf{M}_n} \left(\mathsf{P}_{\mathbf{Y}_{\ell}}(y)-\E{\mathsf{P}_{\mathbf{Y}_{\ell}}(y)}\right)\Biggr| +\mathsf{T}_n
\end{IEEEeqnarray}
where the first inequality follows from the triangle inequality, and the second inequality follows from the fact that $\pi(y),\mu(y)\in[0,1]$. We thus obtain from Hoeffding's inequality~\cite{Hoeffding63} that
\begin{IEEEeqnarray}{lCl}
    \IEEEeqnarraymulticol{3}{l}{P_{[\mathsf{T}_n]}(\hat{\pi}\notin \set{B}_{\varepsilon,\infty}(\pi))} \nonumber\\
  \quad &\leq& \sum_{y\in\set{Y}} P_{[\mathsf{T}_n]}\left(\Biggl|\sum_{\ell=1}^{\mathsf{M}_n} \left(\mathsf{P}_{\mathbf{Y}_{\ell}}(y)-\E{\mathsf{P}_{\mathbf{Y}_{\ell}}(y)}\right)\Biggr| > \varepsilon\mathsf{M}_n - \mathsf{T}_n\right)\nonumber\\
  \quad &\leq& 2|\set{Y}|\exp\left(-2\frac{(\varepsilon\mathsf{M}_n -\mathsf{T}_n)^2}{\mathsf{M}_n}\right). \label{eq:appA-1}
\end{IEEEeqnarray}
Under the assumption that $\mathsf{T}_n = o(\mathsf{M}_n)$, the exponent on the right-hand side of \eqref{eq:appA-1} is equal to $\varepsilon^2\mathsf{M}_n$ asymptotically as $\mathsf{M}_n\to\infty$. Consequently, $P_{[\mathsf{T}_n]}(\hat{\pi}\notin \set{B}_{\varepsilon,\infty}(\pi))$ decays exponentially in $\mathsf{M}_n$ and superexponentially in $n$.

We continue by upper-bounding the error probability $\epsilon(\delta_{\textnormal{mean}})$. Indeed, we have that
\begin{IEEEeqnarray}{lCl}
\epsilon(\delta_{\textnormal{mean}}) & = & \text{Prob}\left(\delta_{\textnormal{mean}}(\mathbf{Y}_1^{\mathsf{M}_n}) \neq [\mathsf{T}_n],\hat{\pi}\in\set{B}_{\varepsilon,\infty}(\pi)\bigr)\right) \nonumber\\
& & {} + \text{Prob}\left(\delta_{\textnormal{mean}}(\mathbf{Y}_1^{\mathsf{M}_n}) \neq [\mathsf{T}_n],\hat{\pi}\notin\set{B}_{\varepsilon,\infty}(\pi)\bigr)\right) \nonumber\\
& \leq & \text{Prob}\left(\delta_{\textnormal{mean}}(\mathbf{Y}_1^{\mathsf{M}_n}) \neq [\mathsf{T}_n],\hat{\pi}\in\set{B}_{\varepsilon,\infty}(\pi)\bigr)\right)\nonumber\\
& & {} + 2|\set{Y}|\exp\left(-2\frac{(\varepsilon\mathsf{M}_n -\mathsf{T}_n)^2}{\mathsf{M}_n}\right) \label{eq:Pablo}
\end{IEEEeqnarray}
where the inequality follows from \eqref{eq:appA-1}. Since $\text{Prob}\bigl(\hat{\pi} \notin \set{B}_{\varepsilon,\infty}(\pi)\bigr)$ decays superexponentially in $n$, the error exponent of $\epsilon(\delta_{\textnormal{mean}})$ is determined by the error exponent of the first term, which we shall bound in the following. 

Indeed, for discrete distributions, relative entropy is continuous. Setting
\begin{equation}\label{eq:mean_test:U}
    U_i(\hat{\pi})\triangleq D\left(\mathsf{P}_{\mathbf{Y}_i} \big\| \hat{\pi}\right)
\end{equation}
it follows that, for $\hat{\pi}\in\set{B}_{\varepsilon,\infty}(\pi)$, $U_i(\hat{\pi})$ lies in $[U_i(\pi)-\varepsilon',U_i(\pi)+\varepsilon']$ for some $\varepsilon'$ that vanishes as $\varepsilon\to 0$. The first term on the RHS of \eqref{eq:Pablo} can thus be upper-bounded as
\begin{IEEEeqnarray}{lCl}
\IEEEeqnarraymulticol{3}{l}{\text{Prob}\left(\delta_{\textnormal{mean}}(\mathbf{Y}_1^{\mathsf{M}_n}) \neq [\mathsf{T}_n],\hat{\pi}\in\set{B}_{\varepsilon,\infty}(\pi)\right)} \nonumber\\
\quad & \le& P_{[\mathsf{T}_n]}\left(\min_{i\leq \mathsf{T}_n} U_i(\pi) \leq \max_{j>\mathsf{T}_n} U_j(\pi)+2\varepsilon'\right) \nonumber\\
 &=&  P_{[\mathsf{T}_n]}\left(\bigcup_{i\le\mathsf{T}_n}\bigcup_{j>\mathsf{T}_n}\left\{ U_i({\pi}) < U_j({\pi})+2\varepsilon' \right\}\right)\nonumber\\
 & \le& \sum_{i=1}^{\mathsf{T}_n}\sum_{j=\mathsf{T}_n+1}^{\mathsf{M}_n} P_{[\mathsf{T}_n]}\left( U_i({\pi}) < U_j({\pi})+2\varepsilon' \right)\nonumber\\
& = & \mathsf{T}_n (\mathsf{M}_n-\mathsf{T}_n) P_{[\mathsf{T}_n]}\left(U_1(\pi) \leq U_{\mathsf{T}_n+1}(\pi)+2\varepsilon'\right) \IEEEeqnarraynumspace \label{eq:Th_mean_1}
\end{IEEEeqnarray}
where the first inequality follows because, for $\hat{\pi}\in\set{B}_{\varepsilon,\infty}(\pi)$, we can approximate $U_i(\hat{\pi})$ by $U_i(\pi) \pm \varepsilon'$, and by then removing the condition $\hat{\pi}\in\set{B}_{\varepsilon,\infty}(\pi)$; the second inequality follows by applying the union bound; and the last equality follows from the symmetry of the problem.

Since $\mathsf{T}_n < \mathsf{M}_n/2$ and, by the theorem's assumption, $\mathsf{M}_n$ is subexponential in $n$, it follows that the error exponent of the RHS of \eqref{eq:Th_mean_1} is determined by the exponent of $P_{[\mathsf{T}_n]}\left(U_1(\pi) \leq U_{\mathsf{T}_n+1}(\pi)+2\varepsilon'\right)$. We thus study this exponent by following the steps in \cite{Li_Universal} for the case where $\pi$ is known. Indeed, we can apply the version of Sanov's theorem given in \cite[Lemma~1]{Li_Universal} to obtain 
\begin{IEEEeqnarray}{lCl}
    \IEEEeqnarraymulticol{3}{l}{- \lim_{n\to\infty} \frac{1}{n} \log P_{[\mathsf{T}_n]}(U_{\mathsf{T}_n+1}(\pi)-U_1(\pi)+2\varepsilon' >0)}\nonumber\\
    \quad & = & \min_{Q_1,Q_2} \{D(Q_1\Vert \mu) + D(Q_2\Vert \pi) \} \label{eq:AppA-LBmin}
\end{IEEEeqnarray}
where the minimum is over all distributions $(Q_1,Q_2)$ satisfying $D(Q_2\Vert\pi)-D(Q_1\Vert\pi)+2\varepsilon'\ge 0$. Using this condition to lower-bound $D(Q_2\Vert\pi)$ by $D(Q_1\Vert\pi)-2\varepsilon'$, and noting that $\varepsilon'$ does not depend on $Q_1$, we can lower-bound the minimum in \eqref{eq:AppA-LBmin} as
\begin{IEEEeqnarray}{lCl}
    \IEEEeqnarraymulticol{3}{l}{\min_{Q_1,Q_2} \left\{D(Q_1\|\mu) + D(Q_2\| \pi)\right\}} \nonumber\\
    \quad &\ge & \min_{Q_1} \left\{D(Q_1\|\mu) + D(Q_1\| \pi)\right\} - 2\varepsilon' \nonumber\\
    & = & 2 B(\mu,\pi) - 2\varepsilon'\label{eq:mean:bound_2}
\end{IEEEeqnarray}
where the minimum in the second line is over all probability distributions $Q_1$ and is equal to $2 B(\mu,\pi)$  \cite[Lemma~2]{Li_Universal}. Combining \eqref{eq:Pablo}--\eqref{eq:mean:bound_2}, we obtain that
\begin{equation}
 \alpha_{\text{mean}} \triangleq \varlimsup_{n\to\infty} -\frac{1}{n} \log \epsilon(\delta_{\textnormal{mean}}) \geq 2 B(\mu,\pi)
\end{equation}
upon letting $\varepsilon'$ tend to zero from above. Since, as noted at the beginning of the proof, $\alpha_{\text{mean}}$ cannot exceed $2B(\mu,\pi)$, Theorem~\ref{thm:mean_test} follows.

\section{Proof of Lemma~\ref{lemma:median_conv}}\label{proof:lemma:median_conv}

Let
\begin{equation}
    m(y)\triangleq \operatorname{median}\{\mathsf{P}_{\mathbf{Y}_1}(y),\dots,\mathsf{P}_{\mathbf{Y}_{\mathsf{M}_n}}(y)\}
\end{equation}
so that
\begin{equation}
    \overline{\pi}(y)=\frac{m(y)}{\sum_{y'\in\set{Y}} m(y')}.
\end{equation}
It can be shown that, if $m(y) \in [\pi(y)-\varepsilon',\pi(y)+\varepsilon']$, then we also have $\bar{\pi}(y)\in [\pi(y)-\varepsilon,\pi(y)+\varepsilon]$ with $\varepsilon = \varepsilon'(1+|\set{Y}|)$. Indeed, $m(y) \in [\pi(y)-\varepsilon',\pi(y)+\varepsilon']$ implies
\begin{equation}
    1-|\set{Y}| \varepsilon' \le \sum_{y\in\set{Y}}m(y) \le 1+|\set{Y}| \varepsilon'
\end{equation}
and hence
\begin{equation}\label{eq:bounds:pi_bar}
    \frac{\pi(y)-\varepsilon'}{1+|\set{Y}| \varepsilon'} \le \overline{\pi}(y) \le \frac{\pi(y)+\varepsilon'}{1-|\set{Y}| \varepsilon'}.
\end{equation}
Rewriting the lower bound, it can be loosened as
\begin{IEEEeqnarray}{lCl}
    \overline{\pi}(y) & \ge & \frac{\pi(y)-\varepsilon'}{1+|\set{Y}| \varepsilon'} \nonumber\\
    &=& \pi(y)-\varepsilon'\frac{1+|\set{Y}|\pi(y)}{1+|\set{Y}| \varepsilon'}\nonumber\\
    &\ge& \pi(y)-\varepsilon'\frac{1+|\set{Y}|}{1+|\set{Y}| \varepsilon'}\nonumber\\
    & \ge &  \pi(y)-\varepsilon'(1+|\set{Y}|)
\end{IEEEeqnarray}
where the third step follows because $\pi(y)\le 1$, and the fourth step follows from the nonnegativity of $|\set{Y}| \varepsilon'$. Along similar lines, we can loosen the upper bound in~\eqref{eq:bounds:pi_bar} as
\begin{equation}
    \overline{\pi}(y)\le \frac{\pi(y)+\varepsilon'}{1-|\set{Y}| \varepsilon'} \le \pi(y)+\varepsilon'(1+|\set{Y}|).
\end{equation}

We can thus upper-bound 
\begin{IEEEeqnarray}{lCl}
    \IEEEeqnarraymulticol{3}{l}{P_{[\mathsf{T}_n]}(\bar{\pi}\notin \set{B}_{\varepsilon,\infty}(\pi))} \nonumber\\
  \quad &=& P_{[\mathsf{T}_n]}\left(\bigcup_{y\in\set{Y}}\left\{\overline{\pi}(y)\notin [\pi(y)-\varepsilon,\pi(y)+\varepsilon]\right\}\right)\nonumber\\
  \quad &\le& \sum_{y\in\set{Y}} P_{[\mathsf{T}_n]}\left(\overline{\pi}(y)\notin [\pi(y)-\varepsilon,\pi(y)+\varepsilon]\right)\nonumber\\
    \quad & \leq & |\set{Y}| \max_{y\in\set{Y}} P_{[\mathsf{T}_n]}\bigl(m(y)\notin [\pi(y)-\varepsilon',\pi(y)+\varepsilon']\bigr) \IEEEeqnarraynumspace \label{eq:lemma_median_1}
\end{IEEEeqnarray}
where the first inequality is the union bound, and the second inequality follows from the fact that, for every $y\in\set{Y}$, the event $\{\overline{\pi}(y)\notin[\pi(y)-\varepsilon,\pi(y)+\varepsilon]\}$ is contained in the event $\{m(y)\notin[\pi(y)-\varepsilon',\pi(y)+\varepsilon']\}$.

We next show the following property of the median of a set of numbers:
\begin{lemma}\label{lem:medianProperty}
 Let $\set{K}=\{k_1,\dots,k_\mathsf{L}\}$ be a set of $\mathsf{L}$ numbers, and let $\set{K}_\set{S}=\{k_i\}_{i\in\set{S}}$ denote a subset of $\set{K}$ indexed by $\set{S}\subseteq[\mathsf{L}]$. For any $\set{S}$ such that $|\set{S}|>\mathsf{L}/2$, we have 
 \begin{equation}\label{eq:medianProperty}
     \min_{i\in\set{S}} k_i\le \operatorname{median}\{k_1,\dots,k_\mathsf{L}\} \le \max_{i\in\set{S}} k_i.
 \end{equation}
\end{lemma}
\begin{IEEEproof}
    The proof follows from ordering the elements of $\set{K}$ and recognizing that the median is the element of $\set{K}$ with ordered index $(\mathsf{L}+1)/2$ (for odd $\mathsf{L}$) or the average of elements $\mathsf{L}/2$ and $\mathsf{L}/2+1$ (for even $\mathsf{L}$). The inequalities in \eqref{eq:medianProperty} can be violated only if the set $\set{S}$ contains ordered indices that are either all smaller or all larger than the ordered index of the median. However, since the set $\set{S}$ has a cardinality strictly larger than $\mathsf{L}/2$, neither is possible. This completes the proof.
\end{IEEEproof}

Since the number of outlier sequences satisfies $\mathsf{T}_n < \mathsf{M}_n/2$, we can apply Lemma~\ref{lem:medianProperty} by letting $\set{S}$ be the types of the sequences drawn from the typical distribution, i.e., $\mathsf{P}_{\mathbf{Y}_i}(y)$, $i=\mathsf{T}_n+1,\dots,\mathsf{M}_n$. We can thus bound the median $m(y)$ by
\begin{equation}
\min_{i > \mathsf{T}_n} \mathsf{P}_{\mathbf{Y}_i} \leq m(y) \leq \max_{i > \mathsf{T}_n} \mathsf{P}_{\mathbf{Y}_i}.
\end{equation}
It follows from the union bound that
\begin{IEEEeqnarray}{lCl}
\IEEEeqnarraymulticol{3}{l}{P_{[\mathsf{T}_n]}\bigl(m(y)\notin [\pi-\varepsilon',\pi+\varepsilon']\bigr)} \nonumber\\
\quad & \leq & P_{[\mathsf{T}_n]}\left(\min_{i> \mathsf{T}_n} \mathsf{P}_{\mathbf{Y}_i}(y)< \pi(y)-\varepsilon'\right)\nonumber\\
    & & {} + P_{[\mathsf{T}_n]}\left(\max_{i>\mathsf{T}_n} \mathsf{P}_{\mathbf{Y}_i}(y)> \pi(y)+\varepsilon'\right)\nonumber\\
    &=& P_{[\mathsf{T}_n]}\left(\bigcup_{i>\mathsf{T}_n} \left\{\mathsf{P}_{\mathbf{Y}_i}(y)< \pi(y)-\varepsilon'\right\}\right)\nonumber\\
    & & {} + P_{[\mathsf{T}_n]}\left(\bigcup_{i>\mathsf{T}_n} \left\{\mathsf{P}_{\mathbf{Y}_i}(y)> \pi(y)+\varepsilon'\right\}\right)\nonumber\\
    &\le& (\mathsf{M}_n-\mathsf{T}_n)  P_{[\mathsf{T}_n]}(\mathsf{P}_{\mathbf{Y}_{\mathsf{T}_n+1}}(y)< \pi(y)-\varepsilon')\nonumber\\
    & & {} + (\mathsf{M}_n-\mathsf{T}_n)  P_{[\mathsf{T}_n]}(\mathsf{P}_{\mathbf{Y}_{\mathsf{T}_n+1}}(y)> \pi(y)+\varepsilon'). \IEEEeqnarraynumspace
\end{IEEEeqnarray}
Since $n \mathsf{P}_{\mathbf{Y}_{\mathsf{T}_n+1}}(y)$ is a binomial random variable with $n$ attempts and success probability $\pi(y)$, we can use Hoeffding's inequality to show that
\begin{subequations}
\begin{IEEEeqnarray}{lCl}
    P_{[\mathsf{T}_n]}(\mathsf{P}_{\mathbf{Y}_{\mathsf{T}_n+1}}(y)< \pi(y)-\varepsilon') &\le& \exp\left(-2n \varepsilon'^2\right)\\
    P_{[\mathsf{T}_n]}(\mathsf{P}_{\mathbf{Y}_{\mathsf{T}_n+1}}(y)> \pi(y)+\varepsilon') &\le& \exp\left(-2n \varepsilon'^2\right).
\end{IEEEeqnarray}
\end{subequations}
Consequently,
\begin{IEEEeqnarray}{lCl}
    \IEEEeqnarraymulticol{3}{l}{P_{[\mathsf{T}_n]}\bigl(m(y)\notin [\pi-\varepsilon',\pi+\varepsilon']\bigr)} \nonumber\\
    \quad & \le & 2(\mathsf{M}_n-\mathsf{T}_n)\exp\left(-2n \varepsilon'^2\right)
\end{IEEEeqnarray}
which together with \eqref{eq:lemma_median_1} proves the lemma.\hspace{1em plus 1fill}\IEEEQEDhere

\section{Proof of Theorem~\ref{thm:median_test}}\label{proof:thm:median_test}

Since the typical error exponent in the universal setting cannot be larger than the error exponent in the setting where $\pi$ and $\mu$ are known, it suffices to bound the typical error exponent from below.

By the law of total expectation, we write the expected value on the RHS of~\eqref{eq:typical_exp} as
\begin{IEEEeqnarray}{lCl}
\IEEEeqnarraymulticol{3}{l}{\mathsf{E}\left[\log\bigl(\epsilon(\delta_{\textnormal{median}}|\bar{\pi})\bigr)\right]} \nonumber\\
& = & \mathsf{E}[\log\bigl(\epsilon(\delta_{\textnormal{median}}|\bar{\pi})\bigr)\big| \bar{\pi} \in \set{B}_{\varepsilon,\infty}(\pi)] P_{[\mathsf{T}_n]}\bigl(\bar{\pi} \in \set{B}_{\varepsilon,\infty}(\pi)\bigr) \nonumber\\
& & {} + \mathsf{E}[\log\bigl(\epsilon(\delta_{\textnormal{median}}|\bar{\pi})\bigr)\big| \bar{\pi} \notin \set{B}_{\varepsilon,\infty}(\pi)] P_{[\mathsf{T}_n]}\bigl(\bar{\pi} \notin \set{B}_{\varepsilon,\infty}(\pi)\bigr)\nonumber\\
& \leq & \mathsf{E}[\log\bigl(\epsilon(\delta_{\textnormal{median}}|\bar{\pi})\bigr)\big| \bar{\pi} \in \set{B}_{\varepsilon,\infty}(\pi)] P_{[\mathsf{T}_n]}\bigl(\bar{\pi} \in \set{B}_{\varepsilon,\infty}(\pi)\bigr)\IEEEeqnarraynumspace \label{eq:th2_1}
\end{IEEEeqnarray}
where the inequality follows because the error probability $\epsilon(\delta_{\textnormal{median}}|\bar{\pi})$ cannot exceed one. 

We next recall that
\begin{equation}
    \delta_{\textnormal{median}}(\mathbf{Y}_1^{(2)},\ldots,\mathbf{Y}_{\mathsf{M}}^{(2)}|\bar{\pi})= \argmax_{\set{S}\in \bar{\set{S}}} \sum_{j\in\set{S}} U_j(\bar{\pi})
\end{equation}
where
\begin{equation}
    U_i(\bar{\pi}) \triangleq D\Bigl(\mathsf{P}_{\mathbf{Y}_i^{(2)}}\Bigm\| \bar{\pi}\Bigr).
\end{equation}
An error occurs if the test statistic $U_j(\bar{\pi})$ of any typical sequence is larger than the test statistic of any outlier sequence. Using this below, we obtain
\begin{IEEEeqnarray}{lCl}
    \IEEEeqnarraymulticol{3}{l}{\epsilon(\delta_\textnormal{median}|\overline{\pi})}\notag\\
    &=& P_{[\mathsf{T}_n]}\left(\min_{\ell\le\mathsf{T}_n} U_\ell(\bar{\pi}) < \max_{j>\mathsf{T}_n} U_j(\bar{\pi}) \right)\nonumber\\
    &=&  P_{[\mathsf{T}_n]}\left(\bigcup_{\ell\le\mathsf{T}_n}\bigcup_{j>\mathsf{T}_n}\left\{ U_\ell(\bar{\pi}) < U_j(\bar{\pi}) \right\}\right)\nonumber\\
    & \le& \sum_{\ell=1}^{\mathsf{T}_n}\sum_{j=\mathsf{T}_n+1}^{\mathsf{M}_n} P_{[\mathsf{T}_n]}\left( U_\ell(\bar{\pi}) < U_j(\bar{\pi}) \right)\nonumber\\
    &=& \mathsf{T}_n (\mathsf{M}_n-\mathsf{T}_n) P_{[\mathsf{T}_n]}(U_{\mathsf{T}_n+1}(\bar{\pi})-U_1(\bar{\pi})>0)\nonumber\\
    &\le& \mathsf{M}_n^2 P_{[\mathsf{T}_n]}(U_{\mathsf{T}_n+1}(\bar{\pi})-U_1(\bar{\pi})>0) \label{eq:median:bound_1}
\end{IEEEeqnarray}
where the first inequality follows from the union bound, and where the subsequent equation follows from the symmetry of the problem. Since by assumption $\bar{\pi} \in \set{B}_{\varepsilon,\infty}(\pi)$, we obtain
\begin{IEEEeqnarray}{lCl}
    \IEEEeqnarraymulticol{3}{l}{U_{\mathsf{T}_n+1}(\bar{\pi})-U_1(\bar{\pi})}\notag\\
    &=&D(\mathsf{P}_{\mathbf{Y}_{\mathsf{T}_n+1}^{(2)}}\Vert \overline{\pi})-D(\mathsf{P}_{\mathbf{Y}_1^{(2)}}\Vert \overline{\pi})\notag\\
    &=& \sum_{y\in\set{Y}} \mathsf{P}_{\mathbf{Y}_{\mathsf{T}_n+1}^{(2)}}(y)\log\frac{\mathsf{P}_{\mathbf{Y}_{\mathsf{T}_n+1}^{(2)}}(y)}{\overline{\pi}(y)}- \mathsf{P}_{\mathbf{Y}_1^{(2)}}(y)\log\frac{\mathsf{P}_{\mathbf{Y}_1^{(2)}}(y)}{\overline{\pi}(y)}\notag\notag\\
    &\le& \sum_{y\in\set{Y}} \mathsf{P}_{\mathbf{Y}_{\mathsf{T}_n+1}^{(2)}}(y)\log\frac{\mathsf{P}_{\mathbf{Y}_{\mathsf{T}_n+1}^{(2)}}(y)}{\pi(y)-\varepsilon}- \mathsf{P}_{\mathbf{Y}_1^{(2)}}(y)\log\frac{\mathsf{P}_{\mathbf{Y}_1^{(2)}}(y)}{\pi(y)+\varepsilon}\notag\\
    &\le& D(\mathsf{P}_{\mathbf{Y}_{\mathsf{T}_n+1}^{(2)}}\Vert \pi)-D(\mathsf{P}_{\mathbf{Y}_1^{(2)}}\Vert \pi) + \tilde{\varepsilon}\label{eq:def:eps_dash}
\end{IEEEeqnarray}
where
\begin{equation}
 \tilde{\varepsilon} \triangleq   2\frac{\varepsilon}{\pi_{\min}-\varepsilon}.
\end{equation}
In \eqref{eq:def:eps_dash}, the last inequality follows from a first-order Taylor series of the first two terms around $\pi$. Since the RHS of \eqref{eq:def:eps_dash} does not depend on $\bar{\pi}$, and $\mathbf{Y}_1^{(2)},\ldots,\mathbf{Y}_{\mathsf{M}}^{(2)}$ are independent of the estimate $\overline{\pi}$, it follows from \eqref{eq:median:bound_1}, \eqref{eq:def:eps_dash} that
\begin{IEEEeqnarray}{lCl}
\IEEEeqnarraymulticol{3}{l}{\mathsf{E}[\log\bigl(\epsilon(\delta_{\textnormal{median}}|\bar{\pi})\bigr)\big| \bar{\pi} \in \set{B}_{\varepsilon,\infty}(\pi)]} \nonumber\\
\quad & \leq & \log P_{[\mathsf{T}_n]}(D(\mathsf{P}_{\mathbf{Y}_{\mathsf{T}_n+1}^{(2)}}\Vert \pi)-D(\mathsf{P}_{\mathbf{Y}_1^{(2)}}\Vert \pi) + \tilde{\varepsilon}>0) \nonumber\\
& & {} + 2 \log \mathsf{M}_n. \label{eq:def:eps_dash_2}
\end{IEEEeqnarray}

Using that $\mathbf{Y}_1^{(2)},\ldots,\mathbf{Y}_{\mathsf{M}}^{(2)}$ are i.i.d.\ according to the respective typical and outlying distributions, we can apply the version of Sanov's theorem given in~\cite[Lemma~1]{Li_Universal} to obtain that
\begin{IEEEeqnarray}{lCl}
    \IEEEeqnarraymulticol{3}{l}{- \frac{1}{n-\lceil \rho n\rceil} \log P_{[\mathsf{T}_n]}(D(\mathsf{P}_{\mathbf{Y}_{\mathsf{T}_n+1}^{(2)}}\Vert \pi)-D(\mathsf{P}_{\mathbf{Y}_1^{(2)}}\Vert \pi) + \tilde{\varepsilon}>0)} \nonumber\\
    \quad\qquad\qquad\qquad\qquad & \to & \min_{Q_1,Q_2} \{D(Q_1\Vert \mu) + D(Q_2\Vert \pi) \}\label{eq:LB_min}
\end{IEEEeqnarray}
as $n$ tends to infinity. In \eqref{eq:LB_min}, the minimum is over all distributions $(Q_1,Q_2)$ satisfying $D(Q_2\Vert\pi)-D(Q_1\Vert\pi)+\tilde{\varepsilon}\ge 0$. Using similar steps as in \eqref{eq:mean:bound_2}, it follows then that
\begin{IEEEeqnarray}{lCl}
    \min_{Q_1,Q_2} \left\{D(Q_1\|\mu) + D(Q_2\| \pi)\right\} & \geq &2 B(\mu,\pi) - \tilde{\varepsilon} \label{eq:median:bound_2}
\end{IEEEeqnarray}
where the minimum over $Q_1$ is over all probability distributions $Q_1$. Since
\begin{equation}
        \frac{1}{n} = \frac{n-\lceil \rho n\rceil}{n}\frac{1}{n-\lceil \rho n\rceil}
\end{equation}
and since, by assumption $\lim_{n\to\infty}\log\mathsf{M}_n/n=0$, we obtain from \eqref{eq:median:bound_1}--\eqref{eq:median:bound_2} that
\begin{IEEEeqnarray}{lCl}
\IEEEeqnarraymulticol{3}{l}{\varliminf_{n\to\infty} -\frac{1}{n} \mathsf{E}\left[\log\bigl(\epsilon(\delta_{\textnormal{median}}|\bar{\pi})\bigr)\big| \bar{\pi} \in \set{B}_{\varepsilon,\infty}(\pi)\right]} \nonumber\\
\quad & \geq & (1-\rho) \bigl(2B(\mu,\pi)-\tilde{\varepsilon}\bigr) \label{eq:th2_2}
\end{IEEEeqnarray}
(where $\varliminf$ denotes the \emph{limit inferior}). Here, the factor $(1-\rho)$ is the limit of $(n-\lceil\rho n \rceil)/n$ as $n\to\infty$.

Applying Lemma~\ref{lemma:median_conv} to the length-$\lceil \rho n \rceil$ sequences $\mathbf{Y}_1^{(1)},\ldots,\mathbf{Y}_{\mathsf{M}_n}^{(1)}$, we obtain that the probability \mbox{$P_{[\mathsf{T}_n]}\bigl(\bar{\pi} \in \set{B}_{\varepsilon,\infty}(\pi)\bigr)$} tends to one as $n\to\infty$. Hence, the theorem follows from \eqref{eq:th2_1} and \eqref{eq:th2_2} upon letting $\rho$ and $\varepsilon$ tend to zero.\hspace{1em plus 1fill}\IEEEQEDhere

\section*{Acknowledgment}
This work was supported in part by the European Union’s HORIZON Research and Innovation Programme under grant agreement No 101120657, project ENFIELD (European Lighthouse to Manifest Trustworthy and Green AI), by the Spanish Ministerio de Ciencia, Innovación y Universidades under Grant PID2024-159557OB-C21 (MICIU/AEI/10.13039/501100011033 and ERDF/UE), and by the  Comunidad de Madrid under Grant IDEA-CM (TEC-2024/COM-89).

\bibliographystyle{IEEEtran}
\bibliography{references}

\end{document}